%% file: main.tex
\documentclass[1p]{elsarticle}

\usepackage{hyperref}

\journal{Journal of Systems and Software}

\usepackage{soul,color}
\usepackage{hyperref}
\usepackage{booktabs}
\newcommand{\lmttfont}{\fontfamily{lmtt}\selectfont}

\newcommand{\revision}[1]{{\textcolor{black} {#1}}}

\usepackage{graphicx}
\usepackage{caption}
\usepackage{subfig}

\usepackage{pdfcomment}
\usepackage{todonotes}

\usepackage{inconsolata}

\usepackage{color}

\definecolor{pblue}{rgb}{0.13,0.13,1}
\definecolor{pgreen}{rgb}{0,0.5,0}
\definecolor{pred}{rgb}{0.9,0,0}
\definecolor{pgrey}{rgb}{0.46,0.45,0.48}
\definecolor{backcolour}{rgb}{0.95,0.95,0.92}

\usepackage{listings}
\usepackage{url}

\usepackage{threeparttable}

\usepackage{amsmath,amssymb,amsfonts}
\usepackage[super]{nth}
\usepackage{adjustbox}

\newcommand{\heaptool}{\textit{RejHeapAnalyzer}}

\usepackage{lipsum}
\usepackage{float}

\usepackage{tikz}

\usepackage{scalerel}

\definecolor{arylideyellow}{rgb}{0.91, 0.84, 0.42}
\definecolor{blond}{rgb}{0.98, 0.94, 0.75}

\newcommand{\tikztriangleright}[1][red,fill=red]{\scalerel*{\tikz \draw[rounded corners=0.1pt,#1] (0,-2.5pt)--++(0,5pt)--++(-30:5pt)--cycle;}{\triangleright}}

\usepackage[most]{tcolorbox}

\begin{document}

\begin{frontmatter}

\title{Software Micro-Rejuvenation for Android Mobile Systems}


\author{Domenico Cotroneo}
\author{Luigi De Simone*}
\author{Roberto Natella}
\author{Roberto Pietrantuono} 
\author{Stefano Russo}

\address{DIETI - Universit\`a degli Studi di Napoli Federico II, Via Claudio 21, 80125 Napoli, Italy
\vspace{-6pt}}



\cortext[correspondingauthor]{Corresponding author\\Email address: luigi.desimone@unina.it (Luigi De Simone)}


\begin{abstract}
Software aging -- the phenomenon affecting many long-running systems, causing performance degradation or an increasing failure rate over mission time, and eventually leading to failure - is known to affect mobile devices and their operating systems, too. Software rejuvenation -- the technique typically used to counteract aging -- may compromise the user's perception of availability and reliability of the personal device, if applied at a coarse grain, e.g., by restarting applications or, worse, rebooting the entire device.

This article proposes a configurable micro-rejuvenation technique to counteract software aging in Android-based mobile devices, acting at a fine-grained level, namely on in-memory system data structures. 
The technique is engineered in two phases. Before releasing the (customized) Android version, a heap profiling facility is used by the manufacturer's developers to identify potentially bloating data structures in Android services and to instrument their code. After release, an aging detection and rejuvenation service will safely clean up the bloating data structures, with a negligible impact on user perception and device availability, as neither the device nor operating system's processes are restarted.
The results of experiments show the ability of the technique to provide significant gains in aging mobile operating system responsiveness and time to failure.
\end{abstract}


\end{frontmatter}

\input{introduction.tex}

\input{related_work.tex}

\input{proposal.tex}

\input{case_study.tex}

\input{discussion.tex}

\input{conclusion.tex}

\bibliography{bibliography}

\end{document}

%% file: introduction.tex
\section{Introduction}
\label{sec:introduction}

Mobile devices like smartphones and tablets are popular resources to communicate and to access Web and Cloud services, such as mail, data storage, e-commerce, and social networks. 
Quality expectations of users about personal devices include that they do not incur software crashes and data losses \cite{capgemini2017world}, but it is challenging for manufacturers to ensure the dependability of their operating systems (OS), as these are significantly complex and feature-rich, are upgraded at a fast pace, and are heavily customized to differentiate from competitors' products \cite{maji2010characterizing,wu2013impact}.

\textit{Software aging} is a threat to the dependability of long-running software systems \cite{cotroneo2014survey}. 
Aging causes the software to \textit{bloat} (due to the accumulation of stale resources, such as memory objects, locks, files, etc.), gradually reducing responsiveness and performance, and eventually leading to crash. 
Empirical studies found that software aging affects also operating systems of mobile devices, like Android \cite{araujo2013investigative,qiao2016empirical,weng2016analysis,cotroneo2016software,cotroneo2020comprehensive}.
Due to the complexity of mobile OS, and often to constraints on testing and bug fixing for time-to-market pressure, it is not feasible to avoid software bloat. Software aging leads to poor quality perceived by users and affects the popularity of personal devices on the market.

\textit{Software rejuvenation} is a technique to heal a software system from aging \cite{huang1995software}. Rejuvenation actions restore a clean state when performance becomes too degraded, or resources are close to exhaustion \cite{cotroneo2014survey}. An example is the restart of an OS process, to bring it back to its initial aging-free state. Rejuvenation is a cost-effective remedy against software aging, since it mitigates its effects without requiring to track down and to fix its root-cause (the so-called \textit{aging-related bugs}, ARBs  \cite{SAR-HB-Cotroneo2020}) -- a difficult and time-consuming task, as ARBs are elusive and typically manifest themselves only in the long term \cite{Frattini2016}. 
As the system may be unavailable during rejuvenation (e.g., unable to serve requests during a restart), it is important to keep the downtime of rejuvenation at a minimum.

This paper proposes a fine-grained and fast software (micro-)rejuvenation technique for the Android OS, acting on its in-memory data structures. 
The technique identifies bloated data structures that might benefit from rejuvenation, to then selectively and safely clean them up. The technique has almost no impact on user perception since neither the device nor Android system processes are restarted.
The proposal complements past own work, where we presented a configurable software agent for Android, developed in the context of an R\&D project with Huawei Technologies Co., Ltd  \cite{cotroneo2019configurable}. The agent monitors and estimates the state of an Android device, to trigger rejuvenation when appropriate. 
We describe a software architecture that automates the proposed rejuvenation technique, integrating it with the detection agent.
The integration yields a complete aging detection and rejuvenation solution, separating detection and rejuvenation scheduling  (\textit{if} and \textit{when} to rejuvenate) from the rejuvenation actions (\textit{how} to rejuvenate). 
We describe the experimental evaluation, which shows that the solution provides significant gains in terms of responsiveness (app launch time) and availability (time to aging failure), with practically no penalty on device availability.

The rest of the paper is organized as follows. 
Section 2 discusses related work and highlights the original contributions of the proposal. 
Section \ref{solution} presents the Android micro-rejuvenation solution and its design choices.
Section \ref{experiment} describes the experiments and discusses the results.  
Section \ref{conclusion} concludes the paper.

%% file: related_work.tex
\section{Related Work}
\label{related}





Most of the research on software rejuvenation focuses on \textit{when} to perform rejuvenation, i.e., on determining the optimal rejuvenation time; this is achieved either by analytic models (time-based rejuvenation) \cite{qiao2018two, xiang2018new}, or by measurements (inspection-based rejuvenation) \cite{huo2018using, qiao2018empirical, weng2017rejuvenation}. Fewer studies investigate \textit{how} rejuvenation should be performed for the best performance. Our proposal focuses on the latter challenge, namely how the rejuvenation action should be implemented in Android to ensure proper performance from the point of view of user perception at a low cost. 

Rejuvenation techniques fall into two broad categories: \textit{i)} System- or application-level actions, which restore an aging-free state of the whole system or applications, and \textit{ii)} Resource-specific actions, which rejuvenate resources suffering from aging at a fine-grained level. The reader may refer to reference \cite{cotroneo2014survey} for a comprehensive survey.

\textit{System-} or \textit{application-level} rejuvenation techniques (also called \textit{application-generic}  \cite{cotroneo2014survey}) include \textit{OS reboot},  \textit{application restart}, \textit{Virtual Machine Monitor (VMM) and Virtual Machine (VM) restart}, and \textit{cluster failover}. They restart the system or its components or replace them with ``fresh" replicas. This way, the system or component is brought back to an aging-free state. These mechanisms are relatively simple to implement; their main drawback is that they are not optimized for specific aging resources, and may suffer from high downtime or performance slow-down during the rejuvenation time. 

\textit{Resource-specific} techniques are tailored solutions (also called \textit{application-specific} \cite{cotroneo2014survey}), which reduce the service downtime by cleaning specific aging-affected resources. To this aim, they exploit features of the system software architecture, 
or of the kind of resources being managed. Examples in the context of operating systems are represented by kernel tables flushing and file system defragmentation. Resource-specific techniques are a minority of approaches, whereas \textit{component restart} is the most studied - there is for instance a remarkable number of proposals that exploit the multi-process architecture of the Apache Web Server. Other techniques that rejuvenate the system state at a fine-grained level, such as \textit{ad hoc} state checkpointing mechanisms \cite{wang1995checkpointing}, are seldom considered. \revision{Finally, another example is showed in \cite{homchaudhuri2020implementation}, in which the author targets the stack overflow problem in a multi-threaded embedded system, and implements a micro-recovery mechanism that prevents the system from a fatal crash.}
While the difficulty and cost of implementing resource-specific techniques are arguably higher than simple restart, their effectiveness in terms of rejuvenation cost is greater.

A sophisticated technique somehow in between the two categories is \textit{micro-reboot} \cite{candea2004microreboot}. While it is conceived to recover quickly from OS failures, by rebooting specific software modules instead of the whole system, it is suited to counteract aging if used as a proactive technique, by rebooting when aging indicators alert for a severe issue, before failures occur. 
However, micro-reboot demands for individually restartable components, which may be dangerous to restart in Android: the services known experimentally to be more likely to age reside all within the \textit{system server} process \cite{cotroneo2020comprehensive}, which is the core of the system, and its restart would have unforeseeable effects.

Researchers have proposed strategies alternative to rejuvenation to combat software aging. An action proposed by Machida \textit{et al.} is \textit{life extension} \cite{machida2012life}, in which a system (e.g., a virtual machine) is allocated additional memory upon aging detection to temporarily prolong its life. Cleaning the part of the state of a system suffering from aging can also be viewed as a life extension technique since the system is not restarted from scratch. 
Another approach specific for mobile devices is \textit{offloading}: aging can be mitigated by offloading part of the computation to a remote server in the Cloud \cite{wu2015software}. This does not eliminate the aging problem but can mitigate it, assuming that a remote server is available in the architecture and accessible when needed. Both life extension and offloading are not classified as conventional software rejuvenation actions but are alternative remedies. 

As for Android, existing studies focus more on aging detection than on rejuvenation. 
\revision{Qiao \textit{et al.}} \cite{qiao2016empirical} present an empirical study on the manifestations of aging-related bugs in Android, confirming (also through static code analysis) that the issue is present and aging bugs have several manifestation patterns.
 
Araujo \textit{et al.} \cite{araujo2013investigative} propose to identify suitable indicators of software aging in Android applications by a measurements-based approach on common resources, such as CPU and memory utilization. Park \textit{et al.} \cite{park2012automated} propose a technique to detect memory leakage by gathering memory execution information in run-time via Process Control Block (PCB) hooking. The approach is based on three steps: identifying PCB; extracting information on memory leakage by intercepting memory allocations/de-allocations by hook function;  judging memory leakage by analyzing the information extracted. 

Shahriar \textit{et al.} \cite{shahriar2014testing} define some memory leak patterns specific to Android applications, and then use fuzz testing to emulate memory leaks and discover the vulnerabilities. Their goal is testing for robustness against memory leak rather than aging detection. The methodology applied by Hussein \textit{et al.} \cite{hussein2015impact} characterizes energy consumption and performance of an Android device, to correlate the choice of the Garbage Collection (GC) algorithm to the experienced energy consumption. The authors discuss alternative GC designs that extend Dalvik's default, mostly-concurrent, mark-sweep collector with generations, and on-the-fly scanning of thread roots. 

Weng \textit{et al.} \cite{weng2016analysis} show that warm rejuvenation (application restart) is not effective for mitigating Android aging, and propose to use active learning based on random forest to build an Android behavior model to schedule rejuvenation \cite{weng2017rejuvenation}.
Xiang \textit{et al.} argue that traditional models for aging and rejuvenation cannot be applied to mobile devices, as they neglect the patterns of usage behavior and experience specific to them \cite{xiang2019,alonso2013comparative}. 
They propose a model-based approach that considers the typical usage of a mobile device (with frequent switches between active and sleep modes) and use Stochastic Petri Nets to model the behavior and properly trigger rejuvenation during low-usage periods, such as when the device is in sleep mode. 

All these studies on Android are mainly concerned with aging detection and with \textit{when} to perform rejuvenation, rather than with \textit{how} to rejuvenate. 
Indeed, rejuvenation is accomplished merely using process or device restart \cite{araujo2013investigative,weng2016analysis, xiang2018new,weng2017rejuvenation, park2012automated, shahriar2014testing, hussein2015impact,xiang2019}.
Our solution is innovative and unique in the following aspects:

\begin{itemize} 

    \item To the best of our knowledge, this is the first complete resource-specific aging detection and rejuvenation solution for Android, acting selectively on  bloating system data structures; 

\item 
      
The technique avoids unavailability, by rejuvenating few selected data structures, instead of system processes or the whole OS.
It is completely transparent to the end-user, who perceives no aging-related failure yet no downtime at all.

\end{itemize}


%


%% file: proposal.tex


\section{Micro-rejuvenation for Android}
\label{solution}
\subsection{Overview}
Previous studies demonstrated that Android suffers from aging \cite{qiao2016empirical,weng2016analysis,cotroneo2016software}, and
pointed out in particular that the phenomenon affects \emph{system processes}, i.e., privileged user-space processes that provide the services of the Android platform, which are accessed by applications through the Inter-Process Communication API (e.g., Binder). 
Examples are the Activity Manager (which launches applications), the Camera Manager, and the WiFi service. Such important services run as threads within the \emph{system server} process.

As most system processes are written in Java, they rely on automated garbage collection to prevent memory leaks, i.e., to free space used by objects that cannot be referenced in the program anymore. Nevertheless, Java software may be affected by memory bloat, due to objects that are unused but still have references in the program, and thus cannot be de-allocated by the GC. In particular, this issue affects \emph{containers} of data structures. A container is an object that contains references to other objects, and these may be related to Java data structures like lists and sets. The aging problem occurs when the objects' references inside these containers are inserted but never removed \cite{jump2007cork,xu2008precise,Ghanavati2019}. 
The accumulation of unused objects causes the increase of the GC overhead (which consumes CPU to inspect the heap, and slows down the entire process), and eventually the exhaustion of heap memory up to software hang or crash. 
Android system processes incur this problem, revealed by increasing trends of memory consumption and memory management overhead \cite{qiao2016empirical,weng2016analysis,cotroneo2016software}.

The approach we propose leverages dynamic analysis to identify Java containers potentially affected by memory bloat, and that are ``safe'' candidates to an action that flushes their content (i.e., objects). 
Developers may review the containers candidate for rejuvenation, and ultimately decide which containers to automatically instrument (with appropriate tools), for them to be rejuvenated when necessary after product release. The solution avoids a full restart of the OS and system processes, to prevent downtime of system services during rejuvenation. 
Since the solution acts selectively on fine-grained parts of the OS state, it can be considered a form of \emph{micro-rejuvenation} \cite{candea2004microreboot,sundaram2007improving}.

\begin{figure}[h]
    \centering
    \includegraphics[scale=0.28]{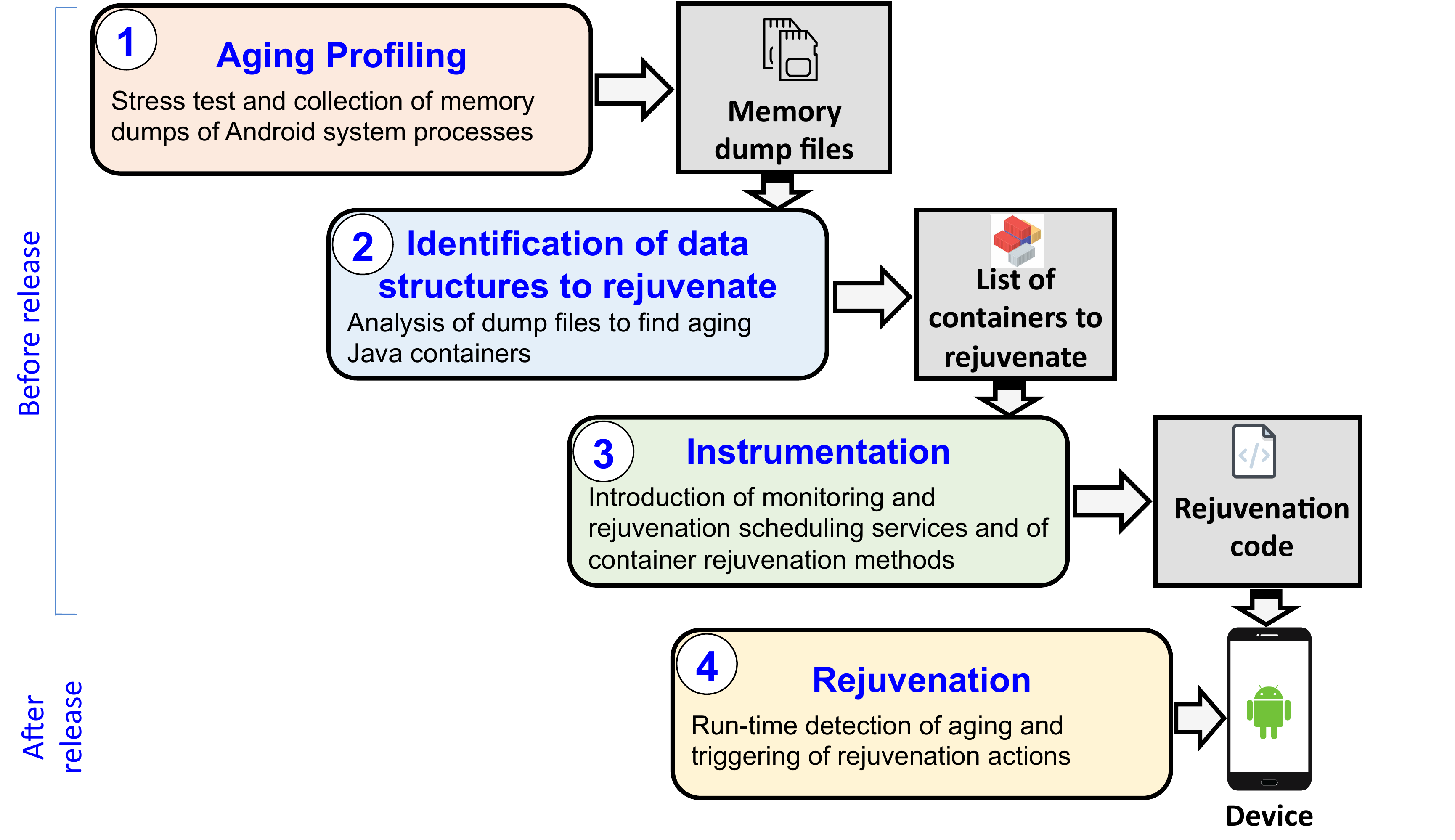}
    \caption{The micro-rejuvenation technique.}
    \label{fig:rejuvenation_workflow}
\end{figure}

The proposed technique consists of the following steps (\figurename{}~\ref{fig:rejuvenation_workflow}):

\begin{enumerate}

\item \textbf{Aging profiling of system processes}: Data on memory consumption of system processes is collected, by generating memory dumps during stress tests (\sectionautorefname{}~\ref{subsec:aging_profiling});

\item \textbf{Identification of containers to rejuvenate}: Memory dumps are processed to identify the Java data structures and the related containers candidate to rejuvenation; 
then, containers which can be safely rejuvenated are selected (\sectionautorefname{}~\ref{subsec:container_selection});

\item \textbf{Instrumentation of system processes}: A rejuvenation routine is introduced into Android system processes, able to clean-up the Java containers identified in the previous step (\subsectionautorefname{}~\ref{subsec:rejuvenation_action});

\item \textbf{Execution of rejuvenation actions}: This is the actual rejuvenation action, that cleans-up Java the selected containers with aging symptoms (\sectionautorefname{}~\ref{subsec:monitoring}).
\end{enumerate}
The first three steps are accomplished by manufacturers once, before product release; the last step automatically rejuvenates an aging Android mobile device before it fails. 

\subsection{Aging profiling}
\label{subsec:aging_profiling}
The first step consists of profiling memory utilization of system processes; gathered data serve in the next step to identify Java containers candidate to micro-rejuvenation. 
Preliminarily, a long-running \emph{stress test} is performed to exercise the OS with an intensive workload as usual in measurements-based aging analyses \cite{SAR-HB-Pietrantuono2020}. 
The test invokes OS services, to make them accumulate objects within Java containers. The subsequent analysis will check if these containers consume resources increasingly over time, which makes them potential sources of memory bloat and candidate to rejuvenation. 

During stress tests, we collect information about memory utilization at the object level, by periodically sampling the entire state of the heap memory of system processes. The state is saved as \textit{memory dump} through {\lmttfont HPROF}, a popular tool for Java heap memory profiling.\footnote{\url{https://docs.oracle.com/javase/8/docs/technotes/samples/hprof.html}} 
Three programs (shell scripts) automate the stress testing and profiling phase:

\begin{itemize}\topsep=0pt

    \item The \textit{application launcher script} starts applications for testing the device. 
    \item The \textit{user workload script} emulates user actions and gestures on the device, leveraging the Android UI/Application Exerciser Monkey tool.\footnote{\label{monkeytool}\url{https://developer.android.com/studio/test/monkey}} We consider the following types of actions/gestures: 
    \begin{itemize}\topsep=0pt
        \item Application switches. The user switches to another application;  
        \item Navigation, which changes the focus from one view to another;
        \item Single touch, when the user taps on the user interface button;
        \item Swipe, when the user moves his/her finger horizontally from right to left;
        \item Multi-touch, when the user presses together several elements of the interface.
    \end{itemize}


    \item The \textit{heap memory dump script} periodically dumps heap memory of a system process under analysis. It uses the {\lmttfont dumpsys} utility to generate {\lmttfont HPROF} dump files.

\end{itemize}


\subsection{Identification of data structures to rejuvenate}
\label{subsec:container_selection}


In this step, the approach identifies, among the thousands of containers in a Java program (e.g., linked lists, hash tables and maps, arrays, vectors, etc.), those that may benefit from rejuvenation. The approach analyzes the time series of memory utilization of Java containers and looks for the ones that exhibit an increasing trend. 

The proposed solution mitigates the accumulation of objects in Java containers, focusing on those that \textit{i)} accumulate objects no longer used, and \textit{ii)} can be emptied without causing side effects on the Java software. The idea is to rejuvenate by getting rid of objects (within a container) holding \textit{soft} information (i.e., data structures whose content is not essential for the correct behavior of Java software). Examples are cached data (such as bitmaps retrieved from storage), log messages, and lists of WiFi networks and Bluetooth devices in the area, which can be easily regenerated when necessary. 

It is important to note that, in principle, it may impossible to find with certainty objects which will not be used in the future. This makes the problem of memory bloat very difficult to address. 
Therefore, we adopt a set of criteria that screen out most of the Java containers that are unlikely to be suitable candidates for rejuvenation. To this aim, we identify a limited set of container categories for further optional review by developers, who know about the semantics of the objects they contain and can finalize the decision to rejuvenate them.
The following categories are identified:

\begin{itemize}\itemsep0pt

    \item \textbf{Containers referenced inside only one class}. The class that includes the container as a member variable is named \textit{dominator}. The \textit{dominator tree} of a memory heap snapshot is a tree in the object reference graph (the directed graph constituted by outgoing references of objects) induced by the following relation: an object \textit{d} dominates an object \textit{y} if every path in the reference graph from any of the garbage collection roots to \textit{y} must go through \textit{d} \cite{SAR-HB-Natella2020}.
    If a container is “hidden” inside the \textit{dominator}, it may be rejuvenated without affecting the users of the \textit{dominator} class (clearly, in conjunction with additional criteria listed below). We do not empty any container that might have an impact outside the \textit{dominator} class. 
    
    \item \textbf{Containers with growing size}. These cause bloating if they ever keep growing. Oppositely, \textit{fixed} containers may hold information strictly required for the correct behavior of the system (e.g., the list of running apps should not be emptied as the OS might lose references to them). They are identified by exercising the Java software (a simple workload may suffice, such as regression test suites), and checking which containers do not vary over time. The containers that fall in this category exhibit a standard deviation of the used heap memory greater than zero;
    
    \item \textbf{Containers that have at least one object with a long lifetime}. Rejuvenation disregards containers that contain only objects with a short lifetime and cannot accumulate (thus, they are unlikely to cause bloat).
    
    \item \textbf{Containers that have objects not accessed for a long time}. Rejuvenation disregards containers whose objects are frequently accessed, since emptying them may cause side effects on Java applications.
    
    \item \textbf{Containers labeled as \textit{disposable} classes by developers}. In general, we refer to the dispose pattern as primarily used in Java to trigger automatic garbage collection. If developers know which Java objects can be considered ``soft'' (such as logs, processes stats, bitmaps), then the technique can be instructed with a \textit{white-list} to identifying containers that contain objects of disposable classes.
    
    \item \textbf{Containers that do not own instances of classes assumed to be critical for the system}. Developers may provide a \textit{black-list} of Java objects they consider \textit{critical} and not to be rejuvenated; the technique will use the black-list to avoid candidating containers including critical objects. 
    
\end{itemize}

The last two categories can be disregarded by developers to speed up the process of building the set of containers to be rejuvenated. The list of candidate containers is built by inspecting the heap memory dumps generated in the profiling step (\sectionautorefname{}~\ref{subsec:aging_profiling}). 


\revision{We address the risk of rejuvenating containers by a judicious combination of dynamic analysis and developers’ feedback. We introduce the six classes of containers mentioned above to identify containers that are good candidates for micro-rejuvenation, with high confidence. This analysis restricts the containers to a small, conservative subset of the entire set of containers in the system. Then, we let the developers make the final decision on which containers to rejuvenate, based on their domain knowledge about the data held by the containers (e.g., transient data). Since the subset of containers is small, it is a feasible trade-off that prevents reliability risks with limited effort by the developers.}

In order to identify containers to be rejuvenated, we implemented a tool named \heaptool{}, which is part of the proposed rejuvenation solution. \heaptool{} is an application that uses the \textit{Eclipse Memory Analyzer} (EMA) framework to automatically analyze heap dump data, helping to construct the list of containers candidate to rejuvenation, referred to as \textbf{\textit{containers-to-rejuvenate list}}.
The memory image dump consists of a set of Object Relational Model databases. The tool inspects them through queries in a SQL-like format. In particular, \heaptool{} \textit{i)} parses heap dump data; \textit{ii)} issues queries on the memory usage of the service; \textit{iii)} ranks the objects by memory usage.
\heaptool{} accepts a set of {\lmttfont HPROF} files. These are parsed and converted into an in-memory representation using EMA. \tableautorefname{}~\ref{tab:sql_hprof_queries} shows queries issued to find containers belonging to the Java {\lmttfont Collections} framework implementation, i.e., object belonging to the {\lmttfont LinkedList}, {\lmttfont HashMap}, {\lmttfont Vector}, {\lmttfont Hashtable}, {\lmttfont ArrayList} classes.

\begin{table}[t]
\resizebox{\textwidth}{!}{

\begin{tabular}{|l|l|}
\hline
\textbf{Target Class} & \textbf{Query} \\ \hline
LinkedList            & {\lmttfont SELECT toString(l.@displayName), l.@retainedHeapSize, inbounds(l).size(), dominatorof(l), l.size FROM java.util.LinkedList l}  \\
HashTable             & {\lmttfont SELECT toString(l.@displayName), l.@retainedHeapSize, inbounds(l).size(), dominatorof(l), l.size FROM java.util.Hashtable l}   \\
ArrayList             & {\lmttfont SELECT toString(l.@displayName), l.@retainedHeapSize, inbounds(l).size(), dominatorof(l), l.size FROM java.util.ArrayList l}    \\
HashMap               & {\lmttfont SELECT toString(l.@displayName), l.@retainedHeapSize, inbounds(l).size(), dominatorof(l), l.size FROM java.util.HashMap l}      \\
Vector                & {\lmttfont SELECT toString(l.@displayName), l.@retainedHeapSize, inbounds(l).size(), dominatorof(l), l.size FROM java.util.Vector l}  \\\hline
\end{tabular}

}

\caption{SQL queries for processing {\lmttfont HPROF} files in \heaptool{}.}
\label{tab:sql_hprof_queries}

\end{table}
Queries give results in the following format: 

\begin{enumerate}\topsep0pt
    \item[-] The in-memory object \textit{name of the container} and its address;

    \item[-] The amount of \textit{memory retained}, which is the memory owned by the object that cannot be freed by the Android platform;

    \item[-] The \textit{number of references} that the container holds to instances of external classes;

    \item[-] The \textit{dominator} of the container. 

    \item[-] The size of the container, which is the number of objects held by the container.
\end{enumerate}

\heaptool{} allows selecting, in the result of a query, containers eligible for rejuvenation, namely those that fall in the first four categories above.

\begin{figure}[hbt]
    \centering
    \includegraphics[width=\columnwidth]{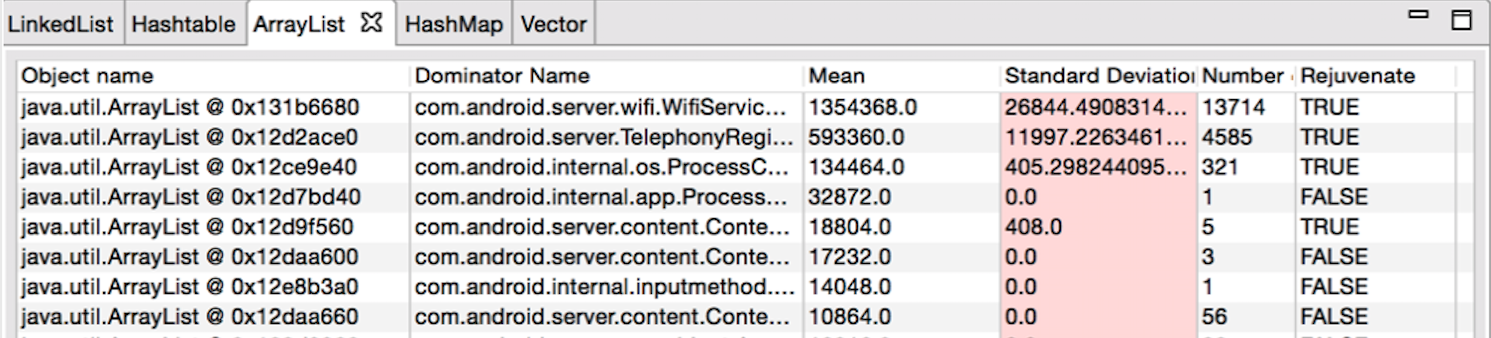}
    \caption{Output of the \heaptool{} tool with ArrayList {\lmttfont system\_server} objects and containers.}
    \label{fig:system_server_containers}
\vspace{-9pt}
\end{figure}

As an example of drastically lowering the number of candidate containers, we issue the queries in \tableautorefname{}~\ref{tab:sql_hprof_queries} targeting the {\lmttfont system\_server} process. The results show that it embeds $12,674$ containers from all Android services. 
The \heaptool{} tool allows reducing the number of candidate containers to $36$. 
\figurename{}~\ref{fig:system_server_containers} shows an excerpt of its output after the analysis of memory dumps of the {\lmttfont system\_server} process; each row lists:
\begin{enumerate}\itemsep0pt

    \item[-] \textbf{Object Name}: the reference of the container;

    \item[-] \textbf{Dominator Name}: the name of the object that dominates the current container;
    
    \item[-] \textbf{Mean}: the expected value of the heap memory in bytes used by that container;
    
    \item[-] \textbf{Standard Deviation}: the standard deviation of the heap memory usage;
    
    \item[-] \textbf{Number}: the number of objects within the container;
    
    \item[-] \textbf{Rejuvenate}: \textit{TRUE}=the container is eligible for rejuvenation, \textit{FALSE} otherwise.

\end{enumerate}


The \textit{containers-to-rejuvenate list} is built by inspecting the output from \heaptool{} and identifying, among all containers of the profiled process, those wasting more memory.
The output of \heaptool{} is a reference set of rejuvenable objects. Each reference has the structure {\lmttfont \textit{TypeOfTheInstance}@\textit{AddressGivenByTheJVM}}. This address structure is the name that the JVM gives to an object, to track its life. The JVM reference name is different from the actual object name in the source code.
In particular, the developer needs to extract details about the \textit{dominator} of the object within the target container. That \textit{dominator} refers indirectly to the Android service to rejuvenate, thus it helps to find the real name of the object to rejuvenate within the service.

\begin{figure}[h]
    \centering
    \includegraphics[scale=0.49]{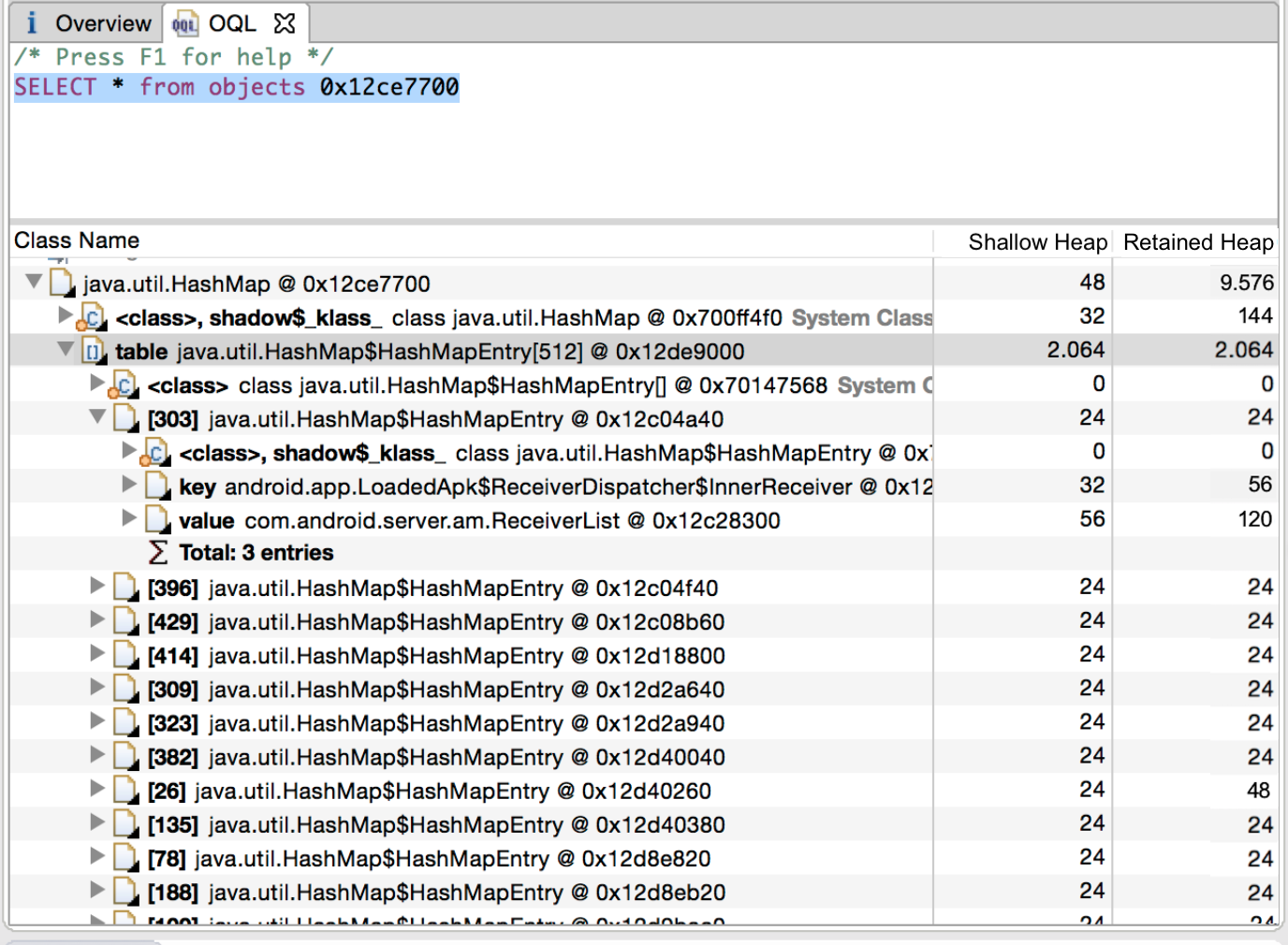}
    \caption{Details of a candidate object in the Eclipse Memory Analyzer framework.}
    \label{fig:object_selection_example}
    
\end{figure}

The developer can extract the object member name querying the EMA framework and the Java Type Inference mechanism. For example, to obtain the information about an object at address {\lmttfont X}, one can submit the query ``{\lmttfont SELECT * from objects X}". 
\revision{\figurename{}~\ref{fig:object_selection_example}}  shows a view of the EMA framework after executing such a query for the object at address {\lmttfont 0x12ce7700}. In the output tree, we notice the node {\lmttfont table}, with the list of objects contained. Every object within a container is identified with a \textit{key-value} pair. In particular, the object referenced by address {\lmttfont 0x12c04a40} belongs to the {\lmttfont ActivityManager} service ({\lmttfont com.android.server.am}), implemented by the {\lmttfont ActivityManagerService} class. 
One can then infer the type of that object ({\lmttfont HashMap<IBinder, ReceiverList>}). Searching for that type in the Android source code, we would find that it is referenced through the member variable  {\lmttfont mRegisteredReceivers}.
Finally, \heaptool{} provides also a filtering feature to select only candidate objects belonging to a given Android component. 

\subsection{Instrumenting for rejuvenation}
\label{subsec:rejuvenation_action}



\begin{figure}[h]
    \centering
    \includegraphics[width=\columnwidth]{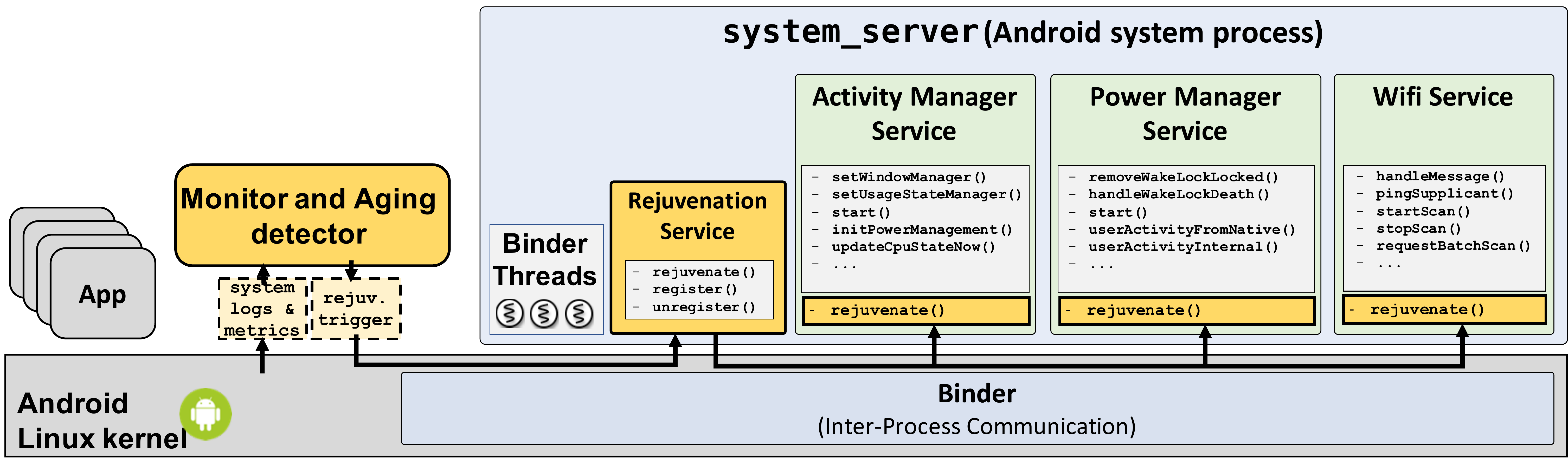}
    \vspace{-7mm}
    \caption{The micro-rejuvenation components within the Android architecture.}
    
    \label{fig:rejuvenation_solution_arch}
\end{figure}

\figureautorefname{}~\ref{fig:rejuvenation_solution_arch} shows the modules of the proposed rejuvenation solution (in yellow) within the Android architecture. Before release, the mobile OS is instrumented with the additional component ``{\lmttfont Monitor and Aging Detector}", and with the automatically generated {\lmttfont RejuvenationService} and {\lmttfont rejuvenate()} procedures to clean the containers in the \textit{containers-to-rejuvenate} list upon aging. 




\revision{Instrumenting source code is a risky process since it might introduce unintended changes to the target software. We minimize this risk by keeping at a minimum the changes to the Android OS, and by making changes at the level of source code (i.e., we refrain from making changes at the bytecode level, which is more prone to reliability issues). Thus, our approach assumes a reliable build system to re-build the Android OS.\\
The proposed approach instruments the services to be rejuvenated, by adding a method that flushes containers ({\lmttfont rejuvenate()}). This method includes up to few tens of lines of code, depending on the containers to be rejuvenated. In our case study, it sufficed to add this method to only three Android services ({\lmttfont ActivityManager}, {\lmttfont PowerManager}, and {\lmttfont WiFi}). Moreover, we instrumented the Android {\lmttfont system\_server} process to introduce a {\lmttfont Binder} interface ($\sim	250$ LOC), in order to listen for requests to trigger rejuvenation, which in turn are dispatched to the services to be rejuvenated.\\
Finally, our approach introduces a monitoring subsystem, which runs as a Java application (few KLOC). Since the monitoring application is non-privileged and runs separately from the Android OS, it does not introduce any additional reliability risk.
}

The {\lmttfont RejuvenationSer\-vice} provides an interface to register Android services that may need to be rejuvenated (\textit{rejuvenable} services), and an interface to trigger rejuvenation. 
To this last aim, the {\lmttfont RejuvenationService} invokes the \texttt{rejuvenate} method (specified in a \texttt{rejuvenable} Java interface) on the registered services. This architecture decouples the rejuvenation action (\textit{how} to rejuvenate) from the aging detection and rejuvenation scheduling (\textit{if} and \textit{when} to rejuvenate). The latter is described in next subsection. 

The {\lmttfont RejuvenationService} is implemented as an Android service within the {\lmttfont system\_\-server} process. It listens for rejuvenation requests over {\lmttfont Binder} (the inter-process communication mechanism for applications and system processes in Android), and dispatches them to the services to rejuvenate, by invoking the rejuvenation method exposed by each \textit{rejuvenable} service. In order to prevent race conditions between rejuvenation and the services, the service temporarily stops requests from apps to the services under rejuvenation, by pausing the Binder threads within the process, using a condition variable.

Every \textit{rejuvenable} service offers a rejuvenation procedure, which cleans 
containers within the \textit{containers-to-rejuvenate list}, containing transient data within the service. 
This code is generated as a method of the service class.
For example, Listing~\ref{lst:rejuvenate_method} shows the method for the {\lmttfont ActivityManager} service. 
The example rejuvenates five containers (note the {\lmttfont mRegisteredReceivers} container discussed in the example in \subsectionautorefname{}~\ref{subsec:container_selection}) that can be flushed without side effects, 
according to tests  performed on the device.

\begin{lstlisting}[caption={Example of {\lmttfont rejuvenate} method},captionpos=b,label={lst:rejuvenate_method},basicstyle=\fontsize{8}{8}\ttfamily]
public void rejuvenate(){
    Slog.i(TAG,"Rejuvenate the Activity Manager Service");
    mPendingProcessChanges.clear();
    mAvailProcessChanges.clear();
    mPendingUidChanges.clear();
    mAvailUidChanges.clear();
    mRegisteredReceivers.clear();
    Slog.i(TAG,"End clean AMS");
}
\end{lstlisting}

The rejuvenation procedure is automatically generated from the \textit{containers-to-rejuve\-nate list} as configured by the developer, and introduced in the service by using Java reflection and code rewriting techniques \cite{frida,xposed}. 
We remark that the \textit{containers-to-rejuvenate list} may be reviewed by developers when customizing the Android release for a specific device. According to their domain knowledge, they may wish to enact more complex solutions, such as removing only the oldest objects from the containers, focusing on objects whose time of last modification is earlier than a given threshold.

\subsection{Run-time rejuvenation actions}
\label{subsec:monitoring}
\begin{figure}[t]
    \centering
        \includegraphics[scale=0.32]{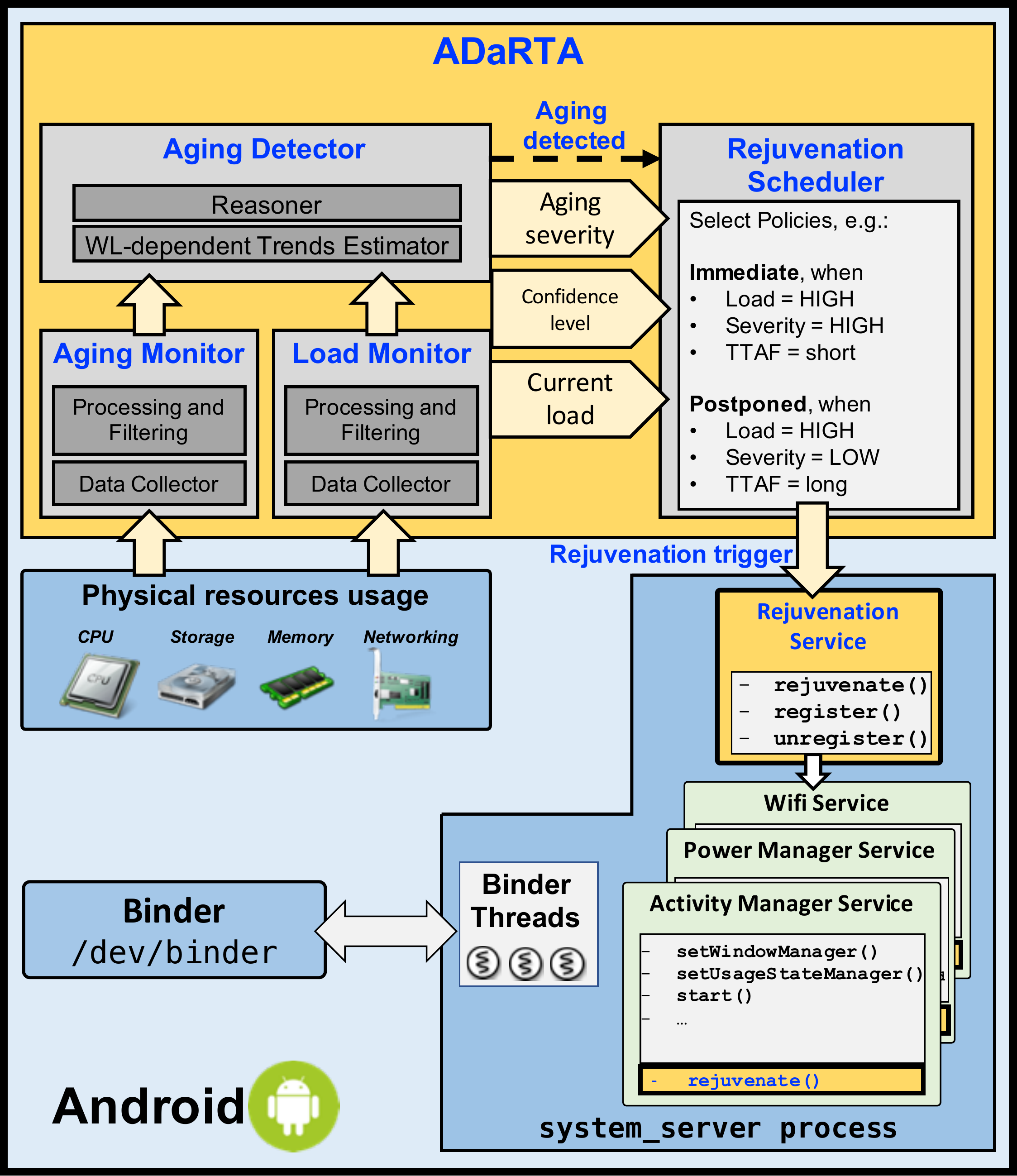}
\vspace{-3pt}
    \caption{Overall aging detection and rejuvenation architecture.}
    \label{fig:adarta_arch}
\end{figure}

The last step actually detects aging and performs rejuvenation actions when the device is in operation. To this aim, as ``{\lmttfont Monitor and Aging Detector}" component (\figurename{}~\ref{fig:rejuvenation_solution_arch}) we use the ADaRTA agent for Android, presented in previous own work \cite{cotroneo2019configurable}. \figurename{}~\ref{fig:adarta_arch} depicts the integration of the proposed micro-rejuvenation solution and the ADaRTA agent. The agent consists of three main modules:
\begin{itemize}\itemsep0pt
    \item \textbf{Aging and Load Monitors}: it collects aging and current workload indicators;

    \item \textbf{Aging Detector}: it analyzes monitoring data, providing an estimate of ongoing aging in terms of severity (Time To Aging Failure, TTAF), confidence, and load level of the device at the time of the detection;
   
    \item \textbf{Rejuvenation Scheduler}: based on the output of the Aging Detector, it decides when to rejuvenate, according to defined policies. Rejuvenation is performed invoking the {\lmttfont RejuvenationService}, as discussed in section~\ref{subsec:rejuvenation_action}.
\end{itemize}

At start-up or after rejuvenation the system is in some \textit{healthy state} (aging free). 
After reboot or rejuvenation, the Detector analyzes the current healthy state to create a reference model, 
consisting of reference thresholds for the \emph{aging indicators} identified in previous work \cite{cotroneo2016software}, such as Launch Time and unused and cache memory. 
The onset of aging is detected by comparing the current state to the healthy state. 

\begin{figure}[H]
    \centering
    \includegraphics[scale=0.41]{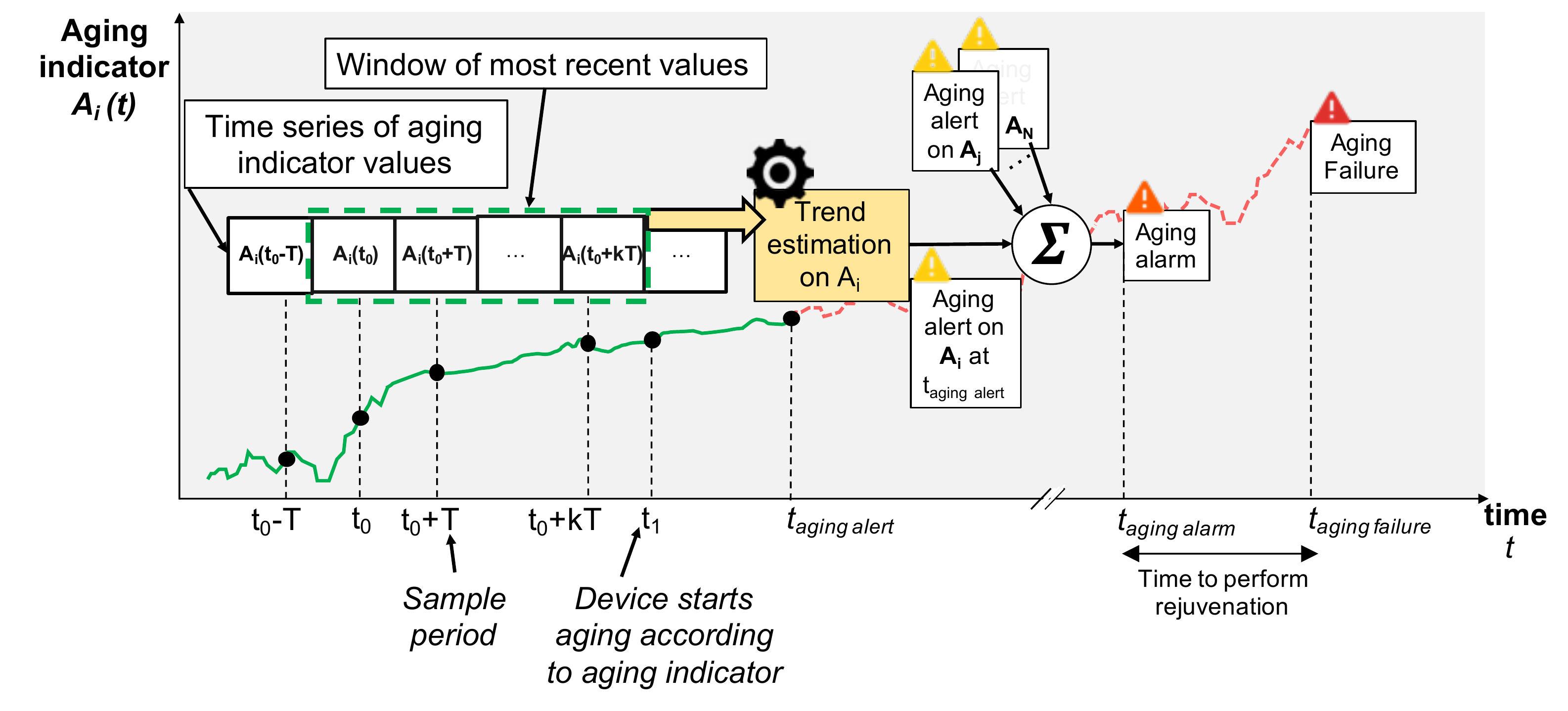}
\vspace{-3pt}
\caption{Aging detection process.}
    \label{fig:time_series_example}
\end{figure}

More specifically, the Detector analyzes \textit{time series} of several aging indicators (produced by the Aging Monitor), to promptly raise an alarm to the Scheduler whenever the device enters an aging state. 
This process is shown in \figurename{}~\ref{fig:time_series_example}. 
Trend estimation techniques -- namely, the \textit{Mann-Kendall} test and the \textit{Sen} procedure - are applied to a sliding-window of samples of every time series. The estimate indicates if the device is affected by performance degradation or resource consumption trends with a defined confidence level (set to 95\% in experiments). 
If a trend persists over a (configurable) period, 
an \textit{aging alert} is raised. Alerts regarding single indicators do not directly trigger rejuvenation: based on the experimental results in \cite{cotroneo2016software}, a set of conditions need to be jointly verified by individual alerts for the Detector to send an actual \textit{aging alarm} to the Scheduler. 
A \textit{confidence level} is assigned to the alarm, according to the set of indicators that concur to it. The higher the confidence level, the stronger the aging phenomenon affecting the device, which is therefore worth to be rejuvenated. 
For instance, two concurrent alerts on both the \textit{Launch Time} and the {\lmttfont system\_server} Proportional Set Size (the portion of RAM occupied by a process) are sufficient to raise an aging alarm with a very high level of confidence. We refer the reader to reference \cite{cotroneo2019configurable} for more details about the used Android aging indicators.

The Scheduler establishes when rejuvenation should be performed, based on the confidence level and the estimated time to failure. The rejuvenation action is actually triggered depending on the current load and on the \emph{policy} configured by the end-user. 
The load is estimated based on measurements of resources usage -- e.g., CPU usage, or number of foreground and background processes - performed by the Load Monitor. 
Configurable policies include: \textit{i)} just warn the user about ongoing aging; \textit{ii)} trigger the remedy action \textit{immediately}; \textit{iii)} \textit{postpone} rejuvenation closer to the expected failure time.

\revision{The \textit{run-time overhead} cost of rejuvenation actions, i.e., the time needed to perform these actions, can be considered negligible. Indeed, the rejuvenation method calls the {\lmttfont clear()} method against containers to be rejuvenated, which simply removes references to the Java objects within the container. The actual deallocation of these objects is delegated to the garbage collector and postponed at a later convenient period. Moreover, since rejuvenation is performed at a very low frequency (one time every several hours), garbage collection does not cause a performance degradation trend.}

%% file: case_study.tex
\section{Experiments}
\label{experiment}

To evaluate the proposed solution in action, we describe five experiments, where we analyze the performance of an Android device for six hours each. We compare the performance of micro-rejuvenating up to three Android services against the case of no rejuvenation. We also compare our approach against full periodic rejuvenation (i.e., device rebooting). The processes and services to rejuvenate are selected based on the empirical evidence reported in the literature \cite{qiao2016empirical,weng2016analysis,cotroneo2016software}, namely that heavy aging effects occur in the {\lmttfont system\_server} process, and in particular in the following Android services running within it: {\lmttfont ActivityManager},  {\lmttfont WiFiService} and {\lmttfont PowerManager}. \revision{We released a public repository, available at this \href{https://github.com/dessertlab/JSSOFTWARE\_Android\_rejuvenation}{link}\footnote{\href{https://github.com/dessertlab/JSSOFTWARE\_Android\_rejuvenation}{https://github.com/dessertlab/JSSOFTWARE\_Android\_rejuvenation}}, with the raw data obtained during the experimental analysis.} The plan includes:

\begin{enumerate}[\textbf{\textit{EXP\arabic{enumi}}.}]

    \item Experiment without rejuvenation;

    \item Experiment rejuvenating only the {\lmttfont ActivityManager} service;
    
    \item Experiment rejuvenating the {\lmttfont ActivityManager} and the {\lmttfont WiFiService} services;

    \item Experiment rejuvenating {\lmttfont ActivityManager}, {\lmttfont WiFiService}, and {\lmttfont PowerManager};
    
    \item Experiment applying full rejuvenation, i.e., rebooting the device periodically.

\end{enumerate}

The experiments are performed on a Google Nexus 5X device, equipped with an Android Open Source Project (AOSP) OS, version 6.0.1. We installed on the device a set of popular mobile apps for exercising it during tests. They include third-party applications downloaded from the Google Play store, identified in a previous experimental work in the collaboration with Huawei \cite{cotroneo2016software}, such as
 {\lmttfont Baidu} (search engine), {\lmttfont Weibo} (social network), {\lmttfont UCMobile} (web browser), {\lmttfont Youku} (video streaming), {\lmttfont Mjweather} (weather).

The workload includes a set of common \textit{events}: application switches, touches, motions, trackballs, and navigation. During the experiments, all these kinds of events are randomly activated by the workload generator, with even probability of occurrence. In order to increase the stress, the applications are \textit{launched} and \textit{killed} every $10$ seconds. We generate these events through the \texttt{adb} shell, via the mentioned Monkey tool.

We collect a sample every time activity within the target application is started (an Android activity encompasses any interaction the user performs \cite{android_activity}). The sample includes the activity \textit{Launch Time} (LT), a symptom of the responsiveness of the device. 
LT is considered the main indicator of the quality of service perceived by users \cite{tanenbaum2014modern,android_startup_time}. Indeed, to improve app responsiveness, developers suggest cold-start a basic application with a responsive GUI within at most $200$ ms \cite{android_app_responsive}. In previous work, we observed that LT is highly correlated to aging \cite{cotroneo2020comprehensive}.

We compute the following metrics at the end of each experiment: 
\begin{itemize}\itemsep1pt
    \item \textbf{The presence of trends in the time series of activities’ \textit{Launch Times}}. 
	We compute this through the Mann-Kendall test, a statistical non-parametric hypothesis test for trend detection. 
    The level of significance expressed by the \textit{p-value} represents the confidence in the result of the test; we set it to the common value of $.05$, meaning that we consider there is a trend in the series if the test has the confidence of $(1-$$p$-$value)\% = 95\%$.
    
    \item \textbf{The trend’s slope of every activity's \textit{Launch Time} time series}. The Mann-Kendall test just tells whether or not a trend is present. 
	The slope (in ms/s) is com\-puted by the non-parametric \textit{Sen’s procedure}. A positive slope in the LT time series means that responsiveness is degrading with $95\%$ of probability, and the degradation is not due to randomness. 
	
	\item \textbf{The increase of \textit{Launch Time}}. This is estimated for every activity based on the computed slopes. For instance, with a slope of $0.003$ ms/s and 6-hour test, the LT is expected to increase by $0.003$ ms/s $\cdot (3600 \cdot 6)$ s $= 64.8$ ms, meaning that the initial LT, whatever it is, increases by $+64.8$ ms after $6$ h. We compute this estimate without rejuvenation (denoted by $LT$) and with-rejuvenation ($LT^r$).  
	
	\item \textbf{The gain in \textit{Launch Time}}. Based on the $LT$ and $LT^r$ estimate, we compute, for every activity, the percentage gain on launch time after 6 hours:
    $$Gain_{LT}\%= \frac{LT-LT^r}{LT} \cdot 100$$
    For instance, a gain of $40\%$ means that the LT after 6h is $40\%$ lower with the proposed technique, compared to no rejuvenation.

	\item \textbf{The gain in the expected time to failure}. We consider a maximum acceptable value for the LT of an activity, which in experiments we set to $200$ ms. 
	We assume that an aging failure occurs whenever the LT exceeds this limit. Based on the estimated slopes $LT$ and $LT^r$, we compute the \textit{expected time to aging failure} without and with rejuvenation ($TTAF$ and $TTAF^r$, respectively). 
	The gain is:

    $$Gain_{TTAF}\%= \frac{TTAF^r-TTAF}{TTAF} \cdot 100$$

    For instance, if for an activity $TTAF$ without rejuvenation was 6 hours and it is extended to $TTAF^r = 8$ hours, then the gain is equal to $2/6 * 100 = 33.33 \%$.
\end{itemize}

\tableautorefname{}~\ref{table:table_rej_activity_manager}, \tableautorefname{}~\ref{table:table_rej_activity_manager_and_wifi}, and \tableautorefname{}~\ref{table:table_rej_activity_manager_and_wifi_and_power} compare the results of experiments with rejuvenation (\textit{EXP2}, \textit{EXP3}, \textit{EXP4}, respectively) to those without rejuvenation (\textit{EXP1}). The tables report the trend’s slope, the LT increase after 6 hours, and the time to achieve the $200$ ms increase in the LT, i.e., the time to a performance failure. We report only applications' activities that are more exercised by the workload, which generate enough samples to perform trend analysis.
The results indicate that all target applications benefit from the rejuvenation, even by rejuvenating only one service ({\lmttfont ActivityManager}). 
In some cases, we observe a strong increase of LT in experiments without rejuvenation compared to those with rejuvenation, in which we notice a strong increase of the expected time to aging failure. 

The average Launch Time gains ($Gain_{LT}$) across all the activities in \textit{EXP2}, \textit{EXP3}, and \textit{EXP4} are $33\%$, $33\%$, $62\%$, respectively. This means that applying rejuvenation on the three services together (\textit{EXP4}) leads to the highest benefits. Practically, after 6 hours of uptime, the proposed solution reduces the LT by $62\%$ in the best case, when the three services are rejuvenated. 

About the \textit{expected time to aging failure gain} ($Gain_{TTAF}$), the averages settle on $53\%$, $86\%$, and $123\%$, for \textit{EXP2}, \textit{EXP3}, and \textit{EXP4} respectively. These values are the percentage increase in the time needed for the occurrence of an aging failure.
Some apps benefit more than others from rejuvenation, depending on the extent of system resources and services they use; these include {\lmttfont Baidu}, {\lmttfont Weibo}, and {\lmttfont UCMobile}.

{
\begin{table}[tp]
\begin{adjustbox}{width=\columnwidth,center}
\centering
\begin{threeparttable}[t]

\begin{tabular}{p{7.5cm} rrr| rrr| cc}
\toprule[1pt]\midrule[0.1pt]
    &  \multicolumn{3}{l}{~~~\textbf{Without rejuvenation}} & \multicolumn{3}{l}{~~~ \textbf{~With rejuvenation}} & \multicolumn{2}{l}{~~~~~~~~\textbf{Gain}} \\
\centering\large\textbf{Activity Class Name}   &    \emph{Slope}   &   \emph{LT increase}   &   \emph{TTAF}    &  \emph{Slope}    &   \emph{LT increase}    &   \emph{TTAF}  &   \emph{Gain$_{LT}$}    &   \emph{Gain$_{TTAF}$}  \\
    &   [ms/s]   &  [ms]   &    [h]   &   [ms/s]   &  [ms]   &    [h]  &   [\%]   &   [\%]  \\
\midrule[1pt]
com.android.packageinstaller.permission.\newline ui.GrantPermissionsActivity   &   0.002   &   53.660   &   22.363   &   0.000   &   0.000   &   +inf   &   +100\%   &   +inf  \\
\midrule
com.baidu.searchbox.MainActivity   &   0.008   &   167.181   &   7.178   &   0.004   &   95.974   &   12.503   &   +43\%   &   +74\%  \\
\midrule
com.moji.mjweather.activity.main \newline .AddCityFirstRunActivity   &   0.009   &   197.860   &   6.065   &   0.006   &   120.111   &   9.991   &   +39\%   &   +65\%  \\
\midrule
com.sina.weibo.SplashActivity   &   0.004   &   95.191   &   12.606   &   0.003   &   57.052   &   21.033   &   +40\%   &   +67\%  \\
\midrule
com.UCMobile.intlcom.uc.browser.InnerUCMobile   &   0.003   &   66.033   &   18.173   &   0.002   &   59.471   &   20.178   &   +10\%   &   +11\%  \\
\midrule
com.youku.phone.ActivityWelcome   &   0.011   &   237.575   &   5.051   &   0.008   &   179.914   &   6.670   &   +24\%   &   +32\%  \\
\midrule
com.youku.phonecom.youku.ui.activity.\newline HomePageActivity   &   0.010   &   225.135   &   5.330   &   0.006   &   132.039   &   9.088   &   +41\%   &   +70\%  \\
\midrule
\textbf{Average}   &      &      &      &      &      &      &   \textbf{+33\%}   &   \textbf{+53\%\tnote{*}}  \\\midrule\bottomrule[1pt]

\end{tabular}

\begin{tablenotes}
     \item[*] The infinite percentage gain (+inf) is excluded from the computation of the mean.
\end{tablenotes}

\end{threeparttable}

\end{adjustbox}

\caption{Performance degradation trends by micro-rejuvenating Java containers of the {\lmttfont ActivityManager} service (\textit{EXP2}).}
\label{table:table_rej_activity_manager}

\end{table}
}

{
\begin{table}[tp]
\begin{adjustbox}{width=\columnwidth,center}

\centering
\begin{tabular}{p{7.5cm} rrr| rrr| cc}
\toprule[1pt]\midrule[0.1pt]
    &  \multicolumn{3}{l}{~~~\textbf{Without rejuvenation}} & \multicolumn{3}{l}{~~~ \textbf{~With rejuvenation}} & \multicolumn{2}{l}{~~~~~~~~\textbf{Gain}} \\
\centering\large\textbf{Activity Class Name}   &    \emph{Slope}   &   \emph{LT increase}   &   \emph{TTAF}    &  \emph{Slope}    &   \emph{LT increase}    &   \emph{TTAF}  &   \emph{Gain$_{LT}$}    &   \emph{Gain$_{TTAF}$}  \\
    &   [ms/s]   &  [ms]   &    [h]   &   [ms/s]   &  [ms]   &    [h]  &   [\%]   &   [\%]  \\
\midrule[1pt]
com.android.packageinstaller \newline .permission.ui.GrantPermissionsActivity   &   0.002   &   53.660   &   22.363   &   0.001   &   15.891   &   75.514   &   +70\%   &   +238\%  \\
\midrule
com.baidu.searchbox.MainActivity   &   0.008   &   167.181   &   7.178   &   0.003   &   74.916   &   16.018   &   +55\%   &   +123\%  \\
\midrule
com.moji.mjweather.activity.main \newline .AddCityFirstRunActivity   &   0.009   &   197.860   &   6.065   &   0.008   &   167.886  &   7.148   &   +15\%   &   +18\%  \\
\midrule
com.sina.weibo.SplashActivity   &   0.004   &   95.191   &   12.606   &   0.001   &   31.689   &   37.868   &   +67\%   &   +200\%  \\
\midrule
com.UCMobile.intlcom.uc.browser.InnerUCMobile   &   0.003   &   66.033   &   18.173   &   0.003   &   60.879  &   19.711   &   +8\%   &   +9\%  \\
\midrule
com.youku.phone.ActivityWelcome   &   0.011   &   237.575   &   5.051   &   0.011   &   244.265   &   4.912   &   -3\%   &   -3\%  \\
\midrule
com.youku.phonecom.youku.ui.activity. \newline HomePageActivity   &   0.010   &   225.135   &   5.330   &   0.009   &   189.597   &   6.329  &   +16\%   &   +19\%  \\
\midrule
\textbf{Average}   &      &      &      &      &      &      &   \textbf{+33\%}   &   \textbf{+86\%}  \\\midrule\bottomrule[1pt]

\end{tabular}
\end{adjustbox}


\caption{Performance degradation trends by micro-rejuvenating Java containers of the {\lmttfont ActivityManager} and {\lmttfont WifiService} services (\textit{EXP3}).}
\label{table:table_rej_activity_manager_and_wifi}

\end{table}
}

{
\begin{table}[tp]
\begin{adjustbox}{width=\columnwidth,center}

\centering
\begin{threeparttable}[tp]

\begin{tabular}{p{7.5cm} rrr| rrr| cc}
  \toprule[1pt]\midrule[0.1pt]
    &  \multicolumn{3}{l}{~~~\textbf{Without rejuvenation}} & \multicolumn{3}{l}{~~~ \textbf{~With rejuvenation}} & \multicolumn{2}{l}{~~~~~~~~\textbf{Gain}} \\
\centering\large\textbf{Activity Class Name}   &    \emph{Slope}   &   \emph{LT increase}   &   \emph{TTAF}    &  \emph{Slope}    &   \emph{LT increase}    &   \emph{TTAF}  &   \emph{Gain$_{LT}$}    &   \emph{Gain$_{TTAF}$}  \\
    &   [ms/s]   &  [ms]   &    [h]   &   [ms/s]   &  [ms]   &    [h]  &   [\%]   &   [\%]  \\
\midrule[1pt]
com.android.packageinstaller \newline .permission.ui.GrantPermissionsActivity   &   0.002   &   53.660   &   22.363   &   0.001   &   12.854   &   93.356   &   +76\%   &   +318\%  \\
\midrule
com.baidu.searchbox.MainActivity   &   0.008   &   167.181   &   7.178   &   -0.004   &   -82,982   &   +inf   &   +150\%   &   +inf\%  \\
\midrule
com.moji.mjweather.activity.main \newline .AddCityFirstRunActivity   &   0.009   &   197.860   &   6.065   &   0.005   &   116.717   &   10.281   &   +41\%   &   +70\%  \\
\midrule
com.sina.weibo.SplashActivity   &   0.004   &   95.191   &   12.606   &   0.002   &   52.076   &   23.043   &   +45\%   &   +83\%  \\
\midrule
com.UCMobile.intlcom.uc.browser.InnerUCMobile   &   0.003   &   66.033   &   18.173   &   0.001   &   25.160   &   47.694   &   +62\%   &   +163\%  \\
\midrule
com.youku.phone.ActivityWelcome   &   0.011   &   237.575   &   5.051   &   0.009   &   201.842   &   5.945   &   +15\%   &   +18\%  \\
\midrule
com.youku.phonecom.youku.ui.activity. \newline HomePageActivity   &   0.010   &   225.135   &   5.330   &   0.006   &   121.723   &   9.858   &   +46\%   &   +85\%  \\
\midrule
\textbf{Average}   &      &      &      &      &      &      &   \textbf{+62\%}   &   \textbf{+123\%\tnote{*}}  \\\midrule[0.2pt]\bottomrule[1pt]
\end{tabular}


\begin{tablenotes}
     \item[*] The infinite percentage gain (+inf) is excluded from the computation of the mean.
\end{tablenotes}

\end{threeparttable}

\end{adjustbox}
\caption{Performance degradation trends by micro-rejuvenating Java containers of the {\lmttfont ActivityManager}, {\lmttfont WifiService}, and {\lmttfont PowerManager} services (\textit{EXP4}).}
\label{table:table_rej_activity_manager_and_wifi_and_power}

\end{table}
}

We point out that performance measurements are subject to sources of non-determin\-ism, due for instance to the garbage collection process. Thus, some results with rejuvenation could be distorted, and, in rare cases, the tests with rejuvenation could exhibit null or worse trends compared to experiments without rejuvenation. In these cases, the rejuvenation did not bring benefit to the activity, and the small negative trend is due to random fluctuations.  
For the activity {\lmttfont com.baidu.searchbox.MainActivity}, in  \textit{EXP4}, we obtain a small negative value of the trend's slope with rejuvenation, hence LT decreases over time. This leads to an ideally infinite $TTAF$. The same applies in \textit{EXP2} to {\lmttfont com.android.packageinstaller.permission.ui.GrantPermissionsActivity}. Despi\-te these few cases, and even taking into account random fluctuations, rejuvenation seems to bring a noticeable improvement for all apps and services.

The performance gains increase with the number of rejuvenated system services (from one to three), although there are diminishing returns when more services are rejuvenated. In some cases, the rejuvenation of {\lmttfont WiFiService} and {\lmttfont PowerManager} services gives comparable gains on average to the case when only the {\lmttfont ActivityManager} is rejuvenated. This is likely since {\lmttfont ActivityManager} is the largest contributor to memory bloating, thus the rejuvenation of the other two services brings less evident improvements. This behavior points out that, from an engineering perspective, it may be sufficient to focus the micro-rejuvenation solution only on the few services that impact the most on software aging. Focusing on a few services simplifies the selection of the containers to rejuvenate, and still brings reasonable improvements.

To set a reference term for the benefit of the proposed technique, in the last experiment (\textit{EXP5}) we perform periodically a full rejuvenation of the device, by rebooting it. This gives an indication of the best achievable gains on activities' launch time and time to failure, since reboot restores a full aging-free system state. The results are in \tableautorefname{}~\ref{table:table_rej_activity_manager_and_wifi_and_power-perfect}. Clearly, periodic reboot provides better LT and TTAF average gains - $75\%$ and $302\%$, respectively - than the proposed technique (Tables~\ref{table:table_rej_activity_manager}-\ref{table:table_rej_activity_manager_and_wifi_and_power}). 
However, a periodic reboot may not be feasible, since a reboot can take few minutes, causing a noticeable impact on device unavailability as perceived by the user. 
As the proposed micro-rejuvenation provides a significant benefit in application responsiveness (launch time) and time-to-aging-failure, without requiring a device restart, we believe it is an appealing solution to Android developers.

\begin{figure}[h]
    \centering
    \includegraphics[width=.5\linewidth]{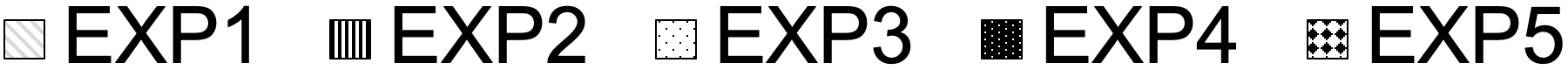}
    \subfloat[Slope]{\label{a}\includegraphics[width=.5\linewidth]{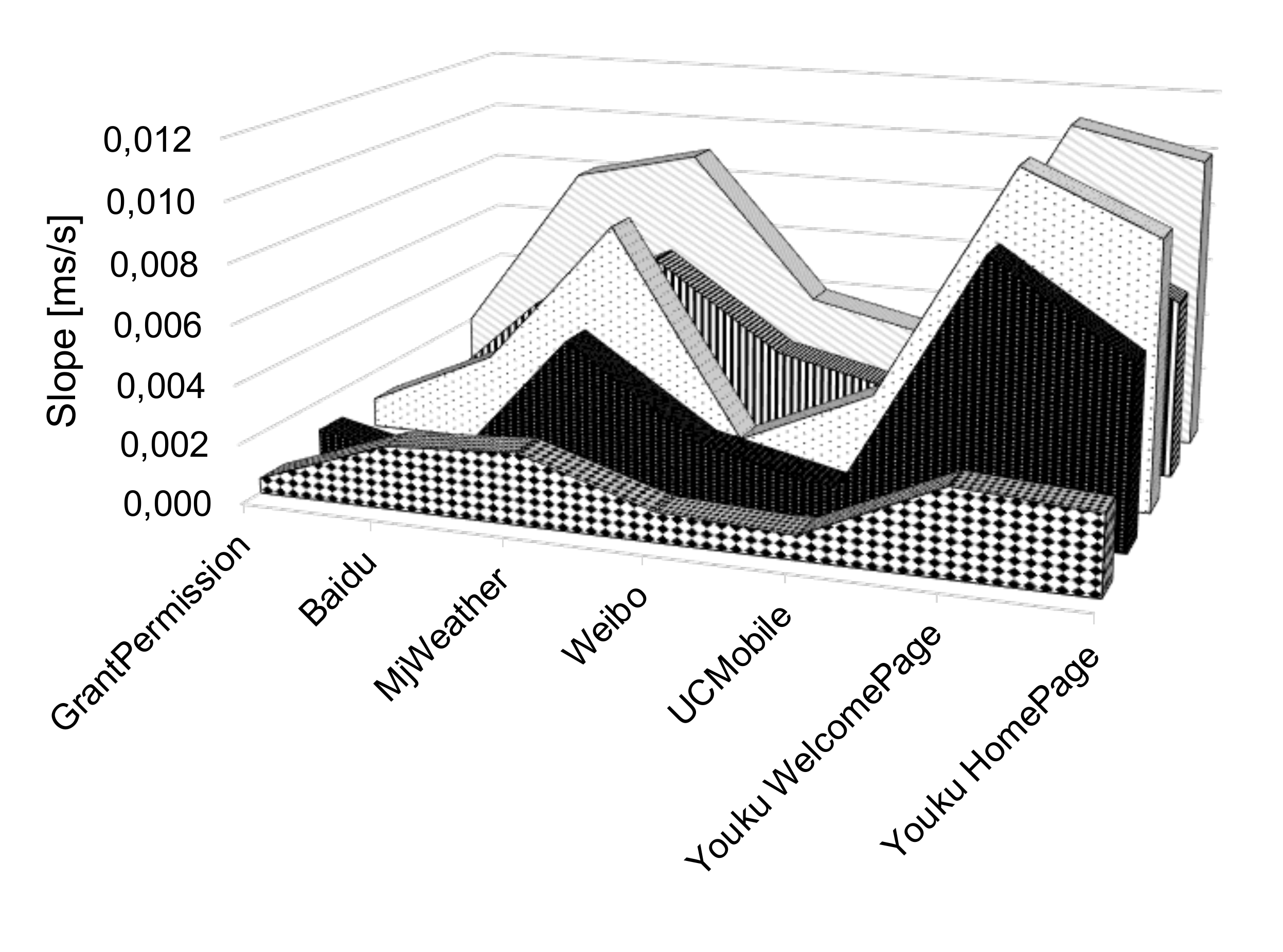}}\hfill
    \subfloat[LT increase]{\label{b}\includegraphics[width=.5\linewidth]{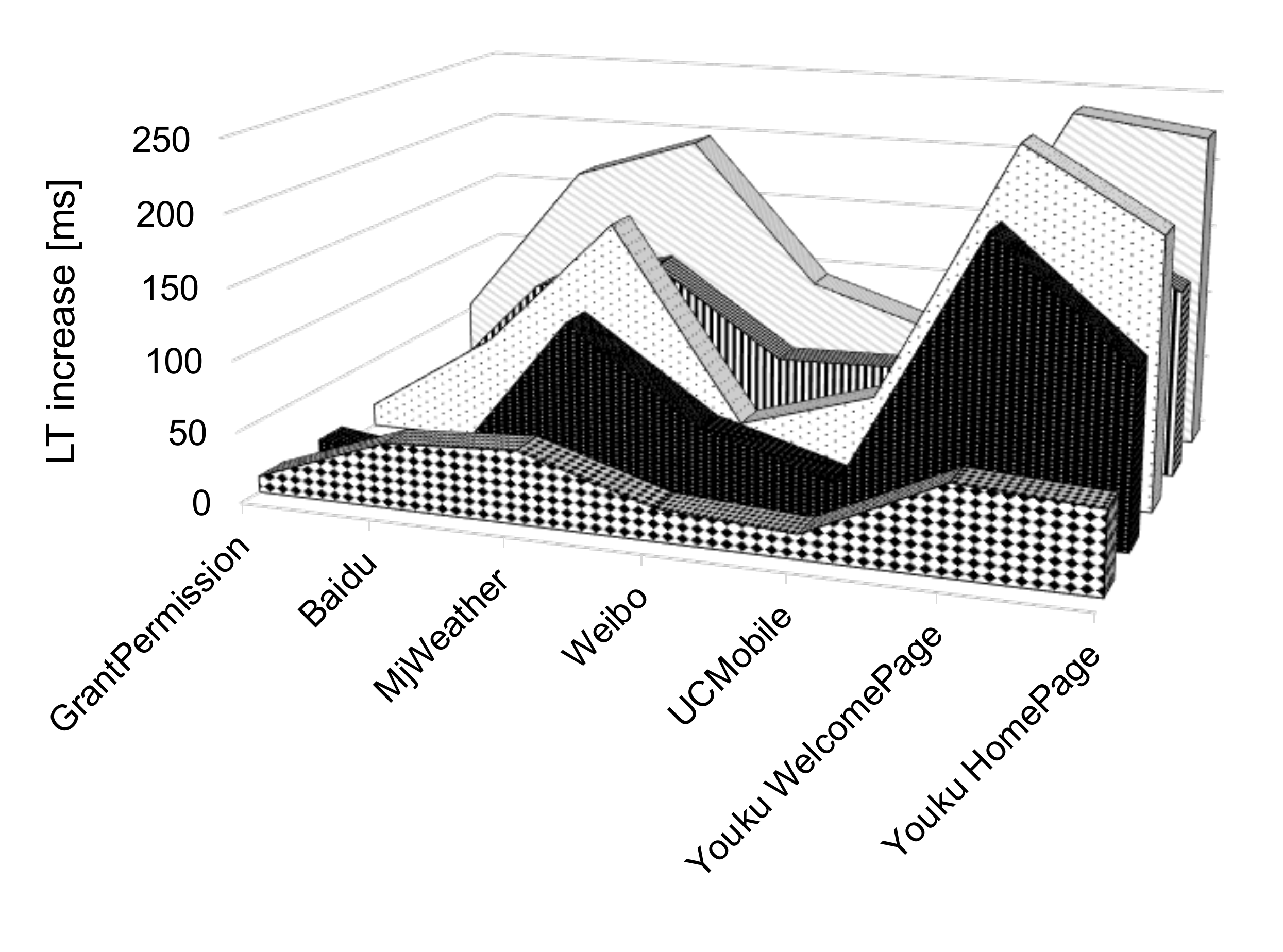}}\par 
    \subfloat[TTAF]{\label{c}\includegraphics[width=.5\linewidth]{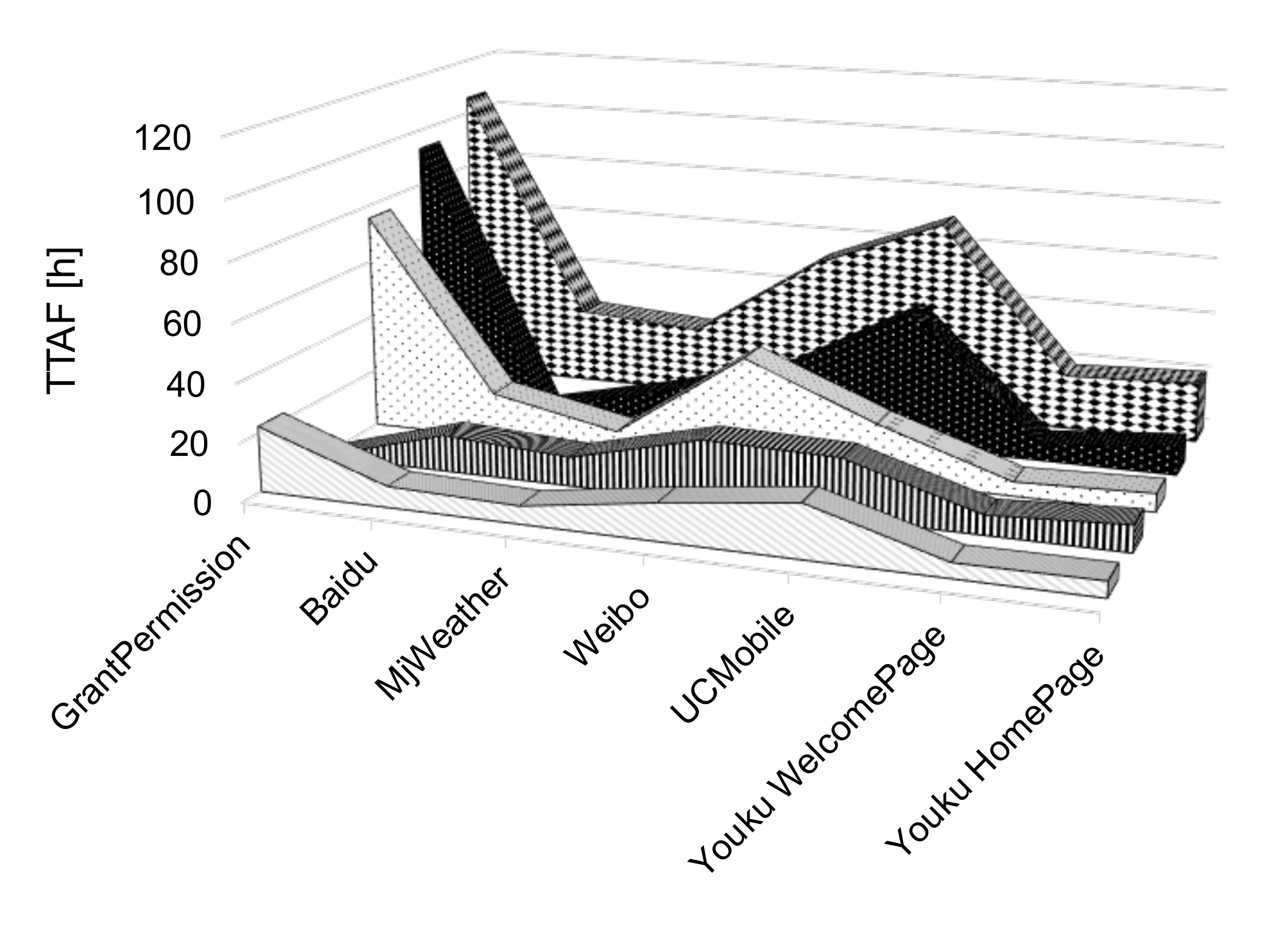}}
    \caption{Trends of \textit{Slope} (lower is better), \textit{LT increase} (lower is better), and \textit{TTAF} (higher is better) for target applications, by varying rejuvenation policies (EXP1 to EXP5).}
    \label{fig:all_policies}
\vspace{-6pt}
\end{figure}

\revision{In addition to the tables discussed above, Figure \ref{fig:all_policies} shows trends of Slope, LT increase, and TTAF according to all rejuvenation policies and target applications. We remark that micro-rejuvenating a few crucial Android services is enough to reach a good trade-off between device availability and time to aging failure.}

{
\begin{table}[t]
\begin{adjustbox}{width=\columnwidth,center}

\centering

\begin{tabular}{p{7.5cm} rrr| rrr| cc}
  \toprule[1pt]\midrule[0.1pt]



    &  \multicolumn{3}{l}{~~\textbf{Without rejuvenation}} & \multicolumn{3}{l}{~~~ \textbf{With rejuvenation}} & \multicolumn{2}{l}{~~~~~~\textbf{Gain}} \\
\centering\large\textbf{Activity Class Name}   &    \emph{Slope}   &   \emph{LT increase}   &   \emph{TTAF}    &  \emph{Slope}    &   \emph{LT increase}    &   \emph{TTAF}  &   \emph{Gain$_{LT}$}    &   \emph{Gain$_{TTAF}$}  \\
    &   [ms/s]   &  [ms]   &    [h]   &   [ms/s]   &  [ms]   &    [h]  &   [\%]   &   [\%]  \\
\midrule[1pt]
com.android.packageinstaller \newline .permission.ui.GrantPermissionsActivity   &   0.002   &   53.660   &   22.363   &   0.0005   &   11.331   &   105.903   &   +79\%   &   +374\%  \\
\midrule
com.baidu.searchbox.MainActivity   &   0.008   &   167.181   &   7.178   &   0.0021   &   45.325   &   26.476   &   +73\%   &   +269\%  \\
\midrule
com.moji.mjweather.activity.main \newline .AddCityFirstRunActivity   &   0.009   &   197.860   &   6.065   &   0.0024   &   50.990   &   23.534   &   +74\%   &   +288\%  \\
\midrule
com.sina.weibo.SplashActivity   &   0.004   &   95.191   &   12.606   &   0.0010   &   22.662   &    52.951   &   +76\%   &   +320\%  \\
\midrule
com.UCMobile.intlcom.uc.browser.InnerUCMobile   &   0.003   &   66.033   &   18.173   &   0.0008   &   16.997   &   70.602   &   +74\%   &   +288\%  \\
\midrule
com.youku.phone.ActivityWelcome   &   0.011   &   237.575   &   5.051   &   0.0029   &   62.321   &   19.255   &   +74\%   &   +281\%  \\
\midrule
com.youku.phonecom.youku.ui.activity \newline HomePageActivity   &   0.010   &   225.135   &   5.330   &   0.0026   &   56.656   &   21.181   &   +75\%   &   +297\%  \\
\midrule
\textbf{Average}   &      &      &      &      &      &      &   \textbf{+75\%}   &   \textbf{+302\%}  \\\midrule[0.3pt]\bottomrule[1pt]
\end{tabular}

\end{adjustbox}
\caption{Performance degradation trends without and with periodic device reboot (EXP5).}
\label{table:table_rej_activity_manager_and_wifi_and_power-perfect}

\end{table}
}



%% file: discussion.tex


\revision{\section{Discussion}}

\noindent
\revision{
$\tikztriangleright[blue,fill=gray!80]$ \textbf{Extending the solution.} It is possible to extend the proposed approach towards mitigating software aging in different operating systems. In our case study, software aging is caused by bloating in Java container classes. Similar issues also occur in other Java-based software systems that execute uninterruptedly for a long time (“long-running”), such as J2EE application servers,e.g., Apache Tomcat. To bring micro-rejuvenation in these systems, developers can apply a similar approach, by identifying Java containers that accumulate disposable data, and by extending the system architecture to safely flush these containers (e.g., by extending the aging classes with a rejuvenation method, and by adding RPC interfaces to trigger rejuvenation).\\\\ 
$\tikztriangleright[blue,fill=gray!80]$ \textbf{Android versions and workload definition.} The location and the extent of software aging in the Android OS can change depending on the version of Android, the applications, and the mobile platform (due to proprietary software customizations made by the vendors). In our work, we focused on the open-source version of Android (AOSP), since it enables us to access the source code and to implement the micro-rejuvenation solution. The sources of software aging that we found (i.e., the three services running within the {\lmttfont system\_server} process) are also used by other Android systems with no significant changes, and represent core services used by all applications, thus they likely also affect many other Android devices with different configurations. Since the problem of memory bloating in object containers is recurrent in Java software, it is reasonable to expect that other Android systems can likely suffer from them, even if on different locations of the Android codebase or to a different extent. In order to apply micro-rejuvenation to a different version with a different workload, the developers can apply our proposed methodology to profile Android services and their Java containers, and identify the ones where to apply rejuvenation.\\\\}

%% file: conclusion.tex
\section{Conclusions}
\label{conclusion}
Mobile devices like smartphones are today probably the largest class of long-running systems -- people tend to rarely switch them off. As such, they suffer from software aging, which manifests itself over long uptime through performance degradation, response time slow down or failures. Many manufacturers equip devices with customized versions of the Android operating system, and, because aging may compromise the user perception of the device quality, they are interested in automatically counteracting it with rejuvenation. 

To this aim, we have presented a fast rejuvenation technique for Android-based mobile devices. The technique acts on in-memory system data structures, providing the advantage of practically no impact on the user perception, as it requires no restart of Android processes or the device. 
The technique supports practitioners in the customization of Android versions, allowing them to add and configure a proactive micro-rejuvenation service to the device operating system. 
We have described how this technique can be engineered through additional Android software modules, able to detect aging trends of system data structures and to automatically rejuvenate them when appropriate. 
We have presented the results of experiments, showing its effectiveness in reducing the degradation of the app launch time and of the time to aging failure of an Android mobile device. 

\revision{
An open research direction for future work is to bring the micro-rejuvenation approach to other types of systems, such as long-running Java-based systems. 
Since the problem of memory bloating in object containers is recurrent in Java software, other systems can also benefit from micro-rejuvenation of Java containers. This approach can provide an additional option for developers, which can achieve a new trade-off between the impact of rejuvenation on availability, and reducing the likelihood of aging-related failures.
}

\vspace{-6pt}